\documentclass[12pt]{article}
\usepackage{amsmath,amssymb,epsfig}
\usepackage{cancel}
\usepackage{color}
\usepackage{caption}

\makeatletter

\@addtoreset{equation}{section}
\makeatother

\begin{document}

\title{\vbox{
\baselineskip 14pt
\hfill \hbox{\normalsize KUNS-2544, EPHOU-15-003,WU-HEP-15-03}
} 
\vskip 1.7cm
\bf D-brane instanton induced $\mu$-terms and their hierarchical structure }
\author{
Hiroyuki~Abe$^{1}$, \ \
Tatsuo~Kobayashi$^{2}$, \ \
Yoshiyuki~Tatsuta$^{1}$, \ \\and \
Shohei~Uemura$^{3}$
\\*[20pt]
{\it \normalsize 
${}^{1}$Department of Physics, Waseda University, 
Tokyo 169-8555, Japan}
\\
{\it \normalsize 
${}^{2}$Department of Physics, Hokkaido University, 
Sapporo, 060-0810 Japan}
\\
{\it \normalsize 
${}^{3}$Department of Physics, Kyoto University, 
Kyoto 606-8502, Japan}
\\*[50pt]}

\date{
\centerline{\small \bf Abstract}
\begin{minipage}{0.9\linewidth}
\medskip 
\medskip 
\small
We study the $\mu$-term matrix of Higgs pairs induced by the D-brane instanton effects in intersecting D6-brane models compactified on $T^6$.
It is found that the $\mu$-term matrix has a certain permutation symmetry and its eigenvalues have large hierarchical structure without fine tuning.
\end{minipage}
}

\newpage

\begin{titlepage}
\maketitle
\thispagestyle{empty}
\clearpage
\end{titlepage}

\renewcommand{\thefootnote}{\arabic{footnote}}
\setcounter{footnote}{0}

\section{Introduction}

The Standard Model is the most successful theory of the particle physics, but some mysteries still remain.
The hierarchy of weak scale and Planck scale is one of them and the supersymmetry may be the solution of it, but supersymmetric models still need another tuning.
In supersymmetric extended standard models, e.g. the minimal supersymmetric standard model (MSSM), 
the Higgs $\mu$-term, $\mu H_u H_d$ needs to be of ${\cal O}(100)$ GeV to realize electro-weak breaking. 
However, there is no reason why $\mu$ is significantly smaller than Planck scale.
In addition to that, it needs to have correct relations to supersymmetry breaking terms, too.
This is known as the ``$\mu$ problem''.

Superstring theory is a promising candidate for unified theory of all the interactions including gravity, and also quarks, leptons and Higgs fields.
Indeed, a number of studies have been done to derive four-dimensional realistic string models through  various compactifications of six dimensions (see for review \cite{Ibanez}).
Among them, intersecting D-brane models are one of interesting model building approaches 
\cite{Bachas:1995ik,Berkooz:1996km,Blumenhagen:2000wh,Aldazabal:2000dg,Angelantonj:2000hi}.(See for a review \cite{Blumenhagen:2006ci}.)
Qualitative aspects of the (supersymmetric) standard model such as 
gauge symmetries, three chiral families and so on  can be realized in some models, e.g. \cite{Ibanez:2001nd,Cvetic:2001nr}.
In these models, $\mu$-terms are often forbidden at the perturbative level by $U(1)$ symmetries, which would be anomalous, 
but could be induced from non-perturbative effects.
The D-brane instanton is a non-perturbative effect in string theory \cite{Blumenhagen:2006xt}.(See also for a review \cite{Blumenhagen:2009qh}.)
The D-brane instanton can break such anomalous $U(1)$  symmetries and produce the $\mu$-terms.
Roughly speaking, such a $\mu$-term is induced at $\mu\sim e^{-S_{cl}}M_s$ where $M_s$ is the string scale and the classical action of the D-brane instanton $S_{cl}$ is almost equal to the volume of the D-brane instanton.
Similarly, right-handed Majorana masses can be induced by D-brane instanton effects \cite{Ibanez:2006da,Ibanez:2007rs,Cvetic:2007ku,Hamada:2014hpa}.

In generic models, there appear more than one Higgs pairs.
For example, intersecting D-brane models with three families on a torus have $\Delta(27)$ flavor symmetry, 
and three families correspond to a triplet of $\Delta(27)$ \cite{Abe:2009vi,BerasaluceGonzalez:2012vb}.\footnote{See also \cite{Higaki:2005ie}.}
Then, triplet Higgs fields can couple with left-handed and right-handed quarks, both of which are 
$\Delta(27)$ triplets.
Actually, in such a toroidal model, there appear, two $\Delta(27)$ triplets of Higgs fields, i.e. 
six pairs of Higgs fields. 
Obviously, multi-Higgs pairs appear in more generic cases, too.

In this paper, we study the $\mu$-term matrix among multi-Higgs pairs induced by the D-brane instanton in toroidal models.
In these models, there remains a permutation symmetry and appears a specific form of the $\mu$-term matrix.
We show that eigenvalues of such a $\mu$-term matrix have a large hierarchy naturally and that could explain the hierarchy between 
very different energy scales, e.g. the huge hierarchy between the  Planck scale and the weak scale.

This paper is organized as follows.
In section 2, we compute the $\mu$-term matrix induced by the D-brane instanton in toroidal models with six pairs of  Higgs fields.
The $\mu$-term matrix depends on the open string moduli of D-branes.
We show some analytic forms of the eigenvalues and eigenvectors of this $\mu$-term matrix.
We generalize this result to models with $n$ pairs of Higgs fields,  too.
In section 3, we compute eigenvalues of the $\mu$-matrix numerically.
We show that the huge hierarchy of eigenvalues, e.g. more than ${\cal O}(10^{10})$, can appear in some parameter region without fine tuning.
Section 4 is devoted to conclusions and discussions.

\section{$\mu$-term in toroidal models}

In this section, we study the $\mu$-term of the intersecting D-brane models.
In most of the intersecting D-brane models, the $\mu$-terms of the Higgs fields are forbidden by (anomalous) $U(1)$ symmetries of D-branes. 
However, such $U(1)$ symmetries  may be broken by the non-perturbative effect.
It was suggested that the D-brane instanton can produce the $\mu$-term\cite{Blumenhagen:2006xt,Blumenhagen:2009qh}.

\subsection{Six-pair models}

For concreteness, we study Type IIA superstring compactified in six dimensional tori $T^6=T^2 \times T^2 \times T^2$ including D6-branes.
In these models, Higgs fields arise from open string NS-sector.
Their zero-modes are localized at the intersection points of D-branes.
We assume that up-type Higgs fields, $H_u$s appear at intersection points between the D6$_a$-brane and D6$_b$-brane, 
and that down-type Higgs fields, $H_d$s appear at intersection points between the D6$_a$-brane and  D6$_c$-brane.
The D6$_a$-brane consists of two parallel branes and has $U(2)$ (or $Sp(2)$) gauge symmetry. 
At first, we study the models having six pairs of Higgs for simplicity, i.e.  $I_{ab}=I_{ac}=6$.
Here, $I_{ab}$ denotes the topological intersection number between D6$_a$-brane and D6$_b$-brane.
This is because models with six pairs of Higgs fields are favored to realize three families of quarks and leptons having Yukawa interactions with Higgs fields perturbatively.
To produce the $\mu$-terms by D2-brane instanton which we name E-brane, 
we need the intersection numbers  $|I_{Ea}|=|I_{Eb}|=|I_{Ec}|=1$.
Zero-modes $\alpha,\beta,\gamma$ appear at each intersection point of E-brane and D6$_{a,b,c}$-brane respectively.
Then, we can calculate the $\mu$-term,
\begin{equation}
\int \mathcal{D}\alpha \mathcal{D}\gamma \mathcal{D}\beta M_s e^{-S_{E}}e^{y_{i}^u  \alpha \cdot H_u^i \beta+y_{j}^d \alpha \cdot H_d^j \gamma}=M_s e^{-S_{E}} y_i^u y_j^d H_u^i H_d^j,
\end{equation}
where $S_E$ is the classical action of E-brane, $M_s$ is the string scale  and  $i=1,\ldots,6$ ($j=1,\ldots,6$) labels the six  $H_u$s ($H_d$s).\footnote{More precisely, there should be a quantum correction factor but we assume it of ${\cal O}(1)$ and ignore it.}
Here, $y_i^u$ is the Yukawa coupling of $\alpha, H_u^i$ and $\beta$, while $y_j^d$ is the Yukawa coupling of $\alpha, H_d^j$ and $\gamma$.

Yukawa couplings in two-dimensional tori $T^2$,  $y_i^u$ and  $y_j^d$,  are generated by world sheet instantons and can be represented by Jacobi $\vartheta$-function \cite{Cvetic:2003ch,Cremades:2004wa,Abe:2009dr}\footnote{
See for a similar calculation in heterotic orbifold models \cite{Hamidi:1986vh}.}
\begin{equation} \vartheta \left[
\begin{array}{c}
a \\
b \\
\end{array}\right] (\nu,\tau) = \sum_{\ell \in {Z}} {\rm exp} \left[ \pi i (a+\ell)^{2} \tau + 2 \pi i (a+\ell)(\nu + b) 
\right].
\label{eq:theta}
\end{equation}
For the $n$-th two-dimensional tori $T_n^2$, we can write
\begin{equation}
y_{i}^{u,n}=\vartheta \left[
\begin{array}{c}
-\frac{i}{I_{ab}^n} +\epsilon^n \\
0  \\
\end{array}\right](0,iI_{ab}^n\mathcal{A}_n/\alpha'),
\end{equation}
\begin{equation}
y_{j}^{d,n}=\vartheta \left[
\begin{array}{c}
-\frac{j}{I_{ac}^n} +\left(\epsilon^n-\frac{\epsilon_c^n}{6} \right) \\
0  \\
\end{array}\right](0,i I_{ac}^n \mathcal{ A}_n/\alpha'),
\end{equation}
where $\mathcal{A}_n$ is the area of the $n$-th tori $T_n^2$ and $I_{ab}^n$ ($I_{ac}^n$) is the intersection number between 
D$6_a$  and D$6_b$ (D$6_a$ and D$6_c$) branes on $T_n^2$ and $\epsilon^n$($\epsilon_c^n$) represents the position of the E-brane (D6$_c$-brane) on $T_n^2$.
These position parameters, $\epsilon^n$ and $\epsilon_c^n$, are normalized such that they vary from 0 to 1.
To realize Yukawa couplings with quarks and leptons perturbatively, 
we set the intersecting numbers, e.g. $(I_{ab,c}^1,I_{ab,c}^2,I_{ab,c}^3)=({6,1,1})$.
Then, we can write 
\begin{equation}
y_i^u=y_{i}^{u,1}y_{}^{u,2}y_{}^{u,3}=\vartheta \left[
\begin{array}{c}
-\frac{i}{6} +\epsilon \\
0  \\
\end{array}\right](0,i6\mathcal{ A}/\alpha')\times C_u,
\end{equation}
\begin{equation}
y_{j}^{d}=y_{j}^{d,1}y_{}^{d,2}y_{}^{d,3}=\vartheta \left[
\begin{array}{c}
-\frac{j}{6} +\left(\epsilon-\frac{\epsilon_c}{6} \right) \\
0  \\
\end{array}\right](0,i 6 \mathcal{A}/\alpha')\times C_d.
\end{equation}
Because the case with $I^2_{ab,c}=I^3_{ab,c}=1$, the $n=2,3$ tori affect only the overall factor of $y_i^u$ and $y_j^d$, here and hereafter, the torus number indices are omitted, and we denote relevant parameters as $\epsilon=\epsilon^1$, $\epsilon_c=\epsilon_c^1$ and ${\cal A}={\cal A}_1$, those determine the form of $\mu$-term matrix.
While the coefficients $C_u$ and $C_d$  determined by parameters of the other tori $n=2,3$ are independent to $i$ and $j$ indices. 
These coefficients would be of ${\cal O}(1)$ and we ignore them hereafter. 
Then, the contribution to $\mu$-term from single E-brane is written as
\begin{equation}
\mu_{ij}= M_s e^{-S_E} \vartheta \left[
\begin{array}{c}
-\frac{i}{6} +\epsilon \\
0  \\
\end{array}\right](0,6i\mathcal{A}/\alpha')
\cdot \vartheta \left[
\begin{array}{c}
-\frac{j}{6} +\left(\epsilon-\frac{\epsilon_c}{6} \right) \\
0  \\
\end{array}\right](0,6i\mathcal{A}/\alpha').
\end{equation}

The E-brane can appear everywhere in the tori.
Thus, we should integrate the E-brane's position parameter $\epsilon$, i.e.,
\begin{equation}
\begin{split}
\mu_{ij} = & M_s e^{-S_E} \int_{0}^{1} d\epsilon \vartheta \left[
\begin{array}{c}
-\frac{i}{6} + \epsilon \\
0  \\
\end{array}
\right] \left( 0,\frac{6i\mathcal{A}}{\alpha'} \right)\cdot
\vartheta \left[
\begin{array}{c}
-\frac{j}{6} + \epsilon - \frac{\epsilon_c}{6}\\
0  \\
\end{array}
\right] \left( 0,\frac{6i\mathcal{A}}{\alpha'} \right),\\
= & M_s e^{-S_E}
\int_{0}^{1} d\epsilon \sum_{m=1}^{2} 
\vartheta \left[
\begin{array}{c}
-\frac{i}{12} - \frac{j}{12} + \epsilon -\frac{\epsilon_c}{12} + \frac{m}{2}\\
0  \\
\end{array}
\right] \left( 0,\frac{12i\mathcal{A}}{\alpha'} \right) \\
&\times \vartheta \left[
\begin{array}{c}
-\frac{i}{12} +\frac{j}{12} +\frac{\epsilon_c}{12}+\frac{m}{2} \\
0  \\
\end{array}
\right] \left( 0,\frac{12i\mathcal{A}}{\alpha'} \right).
\end{split}
\end{equation} 
Then, we obtain 
\begin{equation}
\mu_{ij} =\sqrt{\frac{\alpha'}{12\mathcal{A}}} M_s e^{-S_E} \sum_{m=1}^{2} \vartheta \left[
\begin{array}{c}
-\frac{i-j}{12} + \frac{\epsilon_c}{12}+\frac{m}{2} \\
0  \\
\end{array}
\right] \left( 0,\frac{12i\mathcal{A}}{\alpha'} \right).
\label{eq:mu-term}
\end{equation}
This is the $\mu$-term matrix induced by the D-brane instanton.
This matrix has characteristic aspects.
It is obvious that $\mu_{ij}=\mu_{i'j'}$ if $i-j=i'-j'$.
Similarly, it is found $\mu_{ij}=\mu_{i'j'}$ if $i-j=i'-j'+6$ because
\begin{equation}
\begin{split}
\vartheta \left[
\begin{array}{c}
-\frac{i-j}{12} + \frac{\epsilon_c}{12} \\
0  \\
\end{array}
\right] &\left( 0,\frac{12i\mathcal{A}}{\alpha'} \right)+\vartheta \left[
\begin{array}{c}
-\frac{i-j}{12} + \frac{\epsilon_c}{12} +\frac{1}{2}\\
0  \\
\end{array}
\right] \left( 0,\frac{12i\mathcal{A}}{\alpha'} \right)\\=
&\vartheta \left[
\begin{array}{c}
-\frac{i-j+6}{12} + \frac{\epsilon_c}{12} +\frac{1}{2}\\
0  \\
\end{array}
\right] \left( 0,\frac{12i\mathcal{A}}{\alpha'} \right)+\vartheta \left[
\begin{array}{c}
-\frac{i-j+6}{12} + \frac{\epsilon_c}{12}\\
0  \\
\end{array}
\right] \left( 0,\frac{12i\mathcal{A}}{\alpha'} \right).
\end{split}
\end{equation}
Then, we can summarize general form of the $\mu$-term matrix as follows,
\begin{eqnarray}
\mu_{ij} =\left(
\begin{array}{cccccc}
A & B & C & D & E & F\\
F & A & B & C & D & E\\
E & F & A & B & C & D\\
D & E & F & A & B & C\\
C & D & E & F & A & B\\
B & C & D & E & F & A\\
\end{array}
\right).
\label{eq:Z6-matrix}
\end{eqnarray}
This matrix has a ${\bf Z}_6$ permutation symmetry because the translation symmetry on the torus is discretized by the D-brane's position that restricts the form of the matrix.

The eigenvalues of this matrix $\lambda_{1,\cdots,6}$ are written as follows,
\begin{equation}
\begin{split}
\lambda_1&=A+B+C+D+E+F,\\
\lambda_2&=A+B\omega+C\omega^2 +D\omega^3 +E\omega^4 +F\omega^5,\\
\lambda_3&=A+B\omega^2+C\omega^{4}+D\omega^{6} +E\omega^{8} +F\omega^{10},\\
\lambda_4&=A+B\omega^3+C\omega^{6}+D\omega^{9} +E\omega^{12} +F\omega^{15},\\
\lambda_5&=A+B\omega^4+C\omega^{8}+D\omega^{12} +E\omega^{16} +F\omega^{20},\\
\lambda_6&=A+B\omega^5+C\omega^{10}+D\omega^{15} +E\omega^{20} +F\omega^{25},
\end{split}
\label{list:eigenvalues}
\end{equation}
and their eigenvectors are obtained as 
\begin{equation}
\vec{v}_1=(1,1,1,1,1,1)^t,\vec{v}_2=(1,\omega,\omega^2,\omega^3,\omega^4,\omega^5)^t,\cdots ,\vec{v}_6=(1,\omega^5,\omega^{10},\omega^{15},\omega^{20},\omega^{25})^t,
\label{list:eigenvectors}
\end{equation}
where $\omega = e^{2i\pi/6}$.
In general case, there are no degenerate eigenvalues and the rank of this matrix is 6, but the degeneracies could appear with some special value of $\epsilon_c$.

The first special value of $\epsilon_c$ is 0.
In this case, $y_i^u$ is equal to $y_i^d$ and the $\mu$-term matrix has an extra ${\bf Z_2}$ symmetry.
Then, it becomes the symmetric matrix, 
\begin{eqnarray}
\mu_{ij} =\left(
\begin{array}{cccccc}
A & B & C & D & C & B\\
B & A & B & C & D & C\\
C & B& A & B & C & D\\
D & C & B & A & B & C\\
C & D & C & B & A & B\\
B & C & D & C & B & A\\
\end{array}
\right).
\label{eq:Z2Z6-matrix}
\end{eqnarray}
Then eigenvectors are degenerate as 
\begin{equation}
\begin{split}
\lambda_2=A+B\omega+&C\omega^2+D\omega^3+C\omega^4+B\omega^5\\
&=A+B\omega^{-1}+C\omega^{-2}+D\omega^{-3}+C\omega^{-4}+B\omega^{-5}=\lambda_6,
\end{split}
\end{equation}
and $\lambda_3=\lambda_5$.

The second special value of $\epsilon_c$ is 1/2.
In this case, we can denote $A=B$, $C=F$ and $D=E$.
This is because the D6$_c$-brane lies on the middle point between the $i$-th intersection point and the $(i+1)$-th intersection point.
For example, E2-brane can not distinguish the first and the second pairs of Higgs superfields.
In this case, $\lambda_4=A-A+C-D+D-C=0$.
The $\mu$-term matrix (\ref{eq:mu-term}) has a zero eigenvalue and the rank of (\ref{eq:mu-term}) becomes 5.

\subsection{$n$-pair models}

In the previous section, we computed the $\mu$-term of the six pairs of  Higgs fields.
We can generalize the previous calculation of the $\mu$-term matrix to generic  models with $n$ pairs of Higgs fields.
For $n$ pair models, we only need to generalize the intersection numbers $I_{ab}$ and $I_{ac}$ from $3$ to $n$.
Then, we obtain 
\begin{equation}
\mu_{ij}^{{\rm n-pairs}} \simeq M_s e^{-S_E} \sum_{m=1}^{2} \vartheta \left[
\begin{array}{c}
-\frac{i-j}{2n} + \frac{\epsilon_c}{2n}+\frac{m}{2} \\
0  \\
\end{array}
\right] \left( 0,\frac{2ni\mathcal{A}}{\alpha'} \right),
\label{eq:nmu-term}
\end{equation}
and the ${\bf Z}_{n}$ permutation symmetry appears similarly.
We summarize the $\mu$-term matrix of $n$ pairs as, 
\begin{eqnarray}
\mu_{ij}^{{\rm n-pairs}} =\left(
\begin{array}{ccccc}
A_1 & A_2 & A_3 & \ldots & A_n\\
A_n & A_1 & A_2 & \ldots & A_{n-1}\\
\vdots & \vdots & \vdots & \ddots & \vdots\\
A_2 & A_3 & A_4 & \ldots & A_1\\
\end{array}
\right).
\label{eq:Zn-matrix}
\end{eqnarray}
In these models, $\epsilon_c =0$  and 1/2 are also special values.
The eigenvectors and the eigenvalues are also calculated similarly.

This is the general structure of matricies for D-brane induced quadratic terms in the superpotential.
For example, the right-handed neutrino Majorana mass has the same feature \cite{Hamada:2014hpa}, which can be generated by identifying D6$_c$-brane to D6$_b$-brane and $\epsilon_c$ is vanishing.

\section{Numerical analysis}

\subsection{Six-pair models}
\begin{figure}[thbp]
\begin{tabular}{cc}
\begin{minipage}{0.5\hsize}
\centering
  \epsfig{file=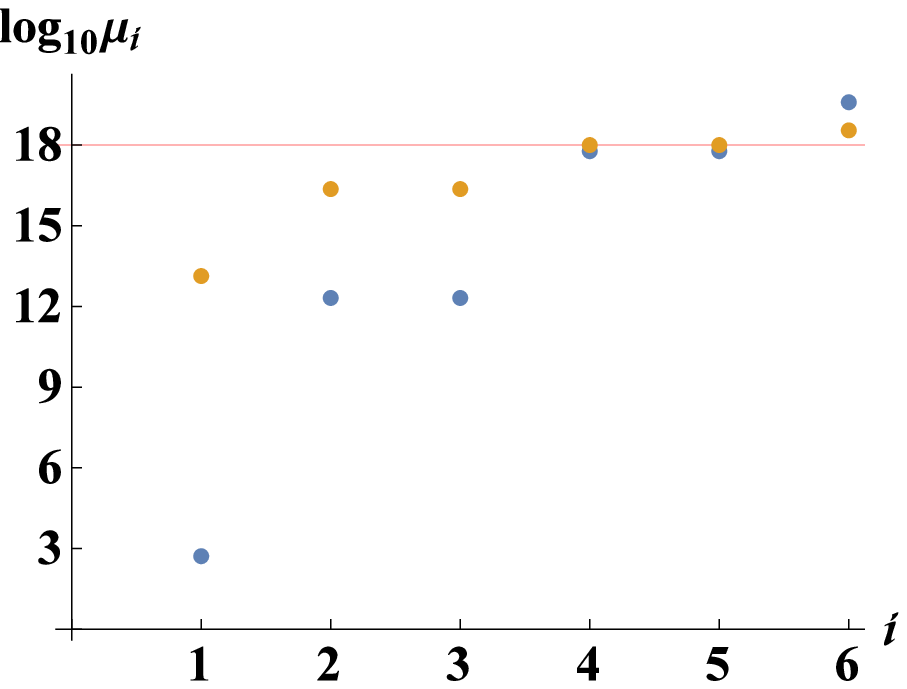,scale=0.8}
 \caption{The eigenvalues of $\mu$-term matrix.The horizontal axis, $i$  denotes the index of the eigenvalues.
We assume $M_se^{-S_E}=10^{18}$GeV. 
The parameters are chosen as 
($12\mathcal{A}/\alpha',\epsilon_c)=(10,0.55)$ for yellow dots and
($12\mathcal{A}/\alpha',\epsilon_c)=(3,0.55)$ for blue dots. \label{fig:graph1}}
\end{minipage}
\hspace{0.2cm}
\begin{minipage}{0.5\hsize}
\centering
  \epsfig{file=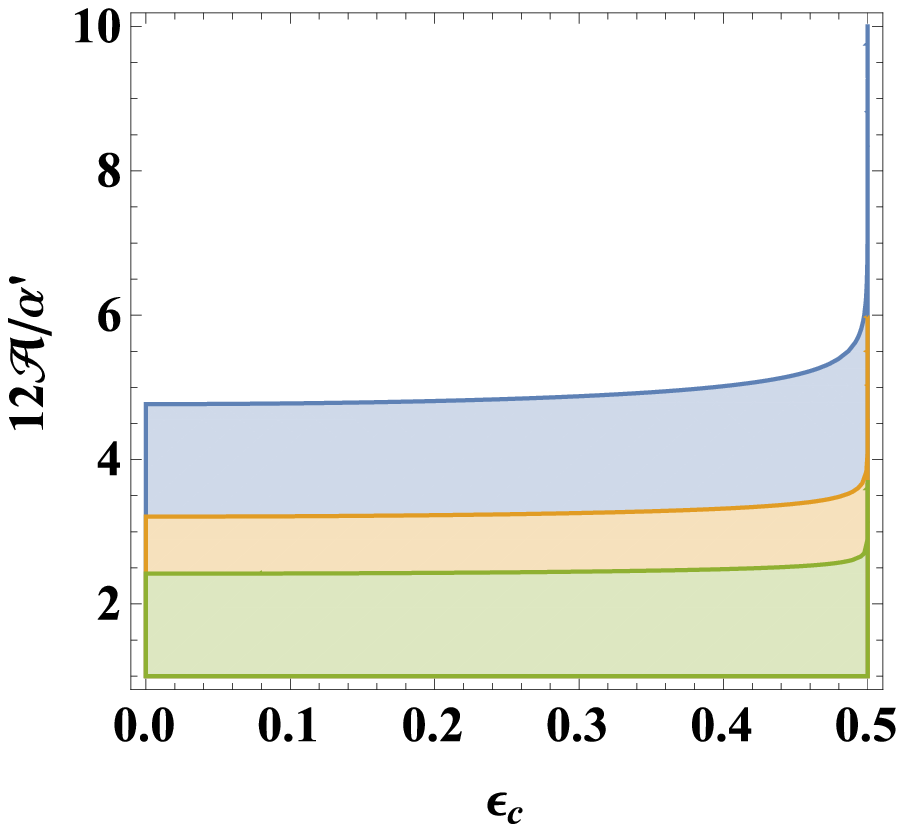,scale=0.8}
 \caption{The blue region denotes the parameter space where $10^{-15}<\lambda_4/\lambda_1<10^{-10}$.
The yellow region denotes $10^{-20}<\lambda_4/\lambda_1<10^{-15}$ and $\lambda_4/\lambda_1$ is smaller than $10^{-20}$ in the green region. \label{fig:graph2}}
\end{minipage}
\end{tabular}
\end{figure}

In this section, we numerically analyze the $\mu$-term matrix (\ref{eq:mu-term}).
Except for overall factors those we omit, the $\mu$-term is determined by the position of the D6$_c$-brane, $\epsilon_c$ and the torus area $\mathcal{A}$.
We show the eigenvalues of (\ref{eq:mu-term}) with some sample values of $\mathcal{A}$ and $\epsilon_c$ in Figure \ref{fig:graph1}.

Figure \ref{fig:graph1} shows that there are one small eigenvalue and not so small five eigenvalues.
The smaller value of $\mathcal{A}$ makes this hierarchy bigger.
This can be understood as follows.
The smallest eigenvalue is $\lambda_4$ in (\ref{list:eigenvalues}), because the $\mu$-term matrix elements have real positive values and  $\omega^3$ is equal to $-1$.
$\lambda_4$ is written as $A-B+C-D+E-F$.
If $\tau$ in (\ref{eq:theta}) is (almost) equal to zero, the theta-function is (almost) independent of $a$ in (\ref{eq:theta}). 
Then, the small torus area $\mathcal{A}$ makes all the $\mu$-term matrix elements almost the same values and the smallest eigenvalue $A-B+C-D+E-F\cong A-A+A-A+A-A=0$.
This cancellation leads to such a hierarchical structure.

Figure \ref{fig:graph2} shows the parameter regions where the hierarchies over $10^{10},10^{15},10^{20}$ appear between the smallest and largest eigenvalues.
This shows if $12\mathcal{A}/\alpha'<5$ or $\epsilon_c\sim 0.5$, the smallest eigenvalue is less than $10^{-10}$ times smaller than the biggest one.
For example, if $M_s e^{-S_E}$ is of ${\cal O}(10^{12})$ GeV, we could realize the weak scale $\mu$-term.
Even if $M_s e^{-S_E}$ is of the order of ${\cal O}(10^{18})$ GeV, we just need $12\mathcal{A}/\alpha'<3$, which means we need no hierarchies among the input parameters to realize the weak scale.

\subsection{$n$-pair models}

We can generalize the previous result to $n$ pair models and have a similar behavior.
We obtain one (or two) hierarchically small eigenvalue(s) and the others are of $\mathcal{O}(M_s e^{-S_{cl}})$ or smaller.
Thus, a huge hierarchy appears.
The number of the smallest eigenstates depends on the number of pairs.
In even pair models, the smallest eigenvalue and the biggest eigenvalue are real and the others form complex conjugate pairs.
In odd pair models, the biggest eigenvalue is the only real eigenvalue and the others (including the smallest eigenvalues) appear as complex conjugate pairs.
This is because the all matrix elements are real.

The absolute value of the smallest eigenvalue in $n$ pair model is shown in Table \ref{tab:even_family} and Table \ref{tab:odd_family}.
\begin{table}[th]
\begin{center}
\begin{tabular}{c|ccc}
\hline
\hline
         & \multicolumn{3}{c}{The lightest eigenvalue$/M_s e^{-S_{cl}}$ }\\
\hline
 number of Higgs pairs & ($2n\mathcal{A}/\alpha',\epsilon_c$)=(5.0,0.55), & (3.0,0.6), & (2.0,0.6) \\
\hline
2 &$10^{-2}$   & $10^{-2}$   & $10^{-3}$\\
4 &$10^{-5}$   & $10^{-7}$   & $10^{-11}$\\
6 &$10^{-10}$ &  $10^{-15}$ & $10^{-24}$\\
8 &$10^{-17}$ &  $10^{-28}$ & $10^{-43}$\\
\hline
\hline
\end{tabular}
\end{center}
\caption{The lightest eigenvalue of the $\mu$-term for even pairs of Higgs fields.
}
\label{tab:even_family}
\end{table}
\begin{table}[h]
\begin{center}
\begin{tabular}{c|ccc}
\hline
\hline
         & \multicolumn{3}{c}{Absolute values of the lightest eigenvalues$/M_s e^{-S_{cl}}$ }\\
\hline
  number of Higgs pairs & ($2n\mathcal{A}/\alpha',\epsilon_c$)=(5.0,0.55), &(3.0,0.6), & (2.0,0.6) \\
\hline
3 &$10^{-1}$   & $10^{-1}$   & $10^{-2}$\\
5 &$10^{-4}$   & $10^{-6}$   & $10^{-10}$\\
7 &$10^{-10}$ &  $10^{-15}$ & $10^{-23}$\\
\hline
\hline
\end{tabular}
\end{center}
\caption{The lightest eigenvalue of the $\mu$-term for odd pairs of  Higgs fields.
}
\label{tab:odd_family}
\end{table}

These tables show more pairs make the smallest eigenvalues smaller.
The reason is understood as follows.
Intersecting D-brane  models are T-dual to magnetized D-brane models.
Thus, in the T-dual picture, 
Yukawa couplings by the worldsheet instanton are equal to the overlap integrals of zero-mode wave functions in magnetized D-brane \cite{Cremades:2004wa}.
If the  number  of Higgs pairs increases, the peaks of these wave functions get closer to each other.
As the wave functions are normalized to 1, the overlap integral gets closer to 1.
This makes the cancellation of matrix elements stronger and the lightest eigenvalue gets closer to 0.
Consequently, we can realize a hierarchy between the smallest and biggest eigenvalues of the $\mu$-term matrix 
in a certain parameter region without fine tuning.

As mentioned before, D-brane instanton effects can also induce the Majorana masses of three families for right-handed neutrino.
The Majorana mass matrix is obtained through a similar calculation by setting $\epsilon_c=0$ \cite{Hamada:2014hpa}.
For example, the case with $n=3$ in  Table \ref{tab:odd_family} does not show a large hierarchy, 
although it corresponds to $\epsilon_c \neq 0$.
The parameter $\epsilon_c=0$ also leads a similar or milder hierarchy for $n=3$.




\section{Conclusion and discussion}

We have studied the D-brane instanton induced $\mu$-terms in toroidal models and carried out their numerical analysis.
The $\mu$-term is determined by the torus area and D-brane position.
It is remarkable that each element of the $\mu$-term matrix is of ${\cal O}(M_se^{-S_E})$, 
but there would appear a huge hierarchy among their eigenvalues without fine tuning.
The smallest eigenvalue could be as small as the weak scale.
Thus, D-brane instanton effects generate a huge hierarchy among 
eigenvalues of the $\mu$-term matrix in a certain parameter space.
Thus, only the lightest Higgs pairs would be important 
for low-energy physics even if there are many pairs below the string scale.
Our analysis would be extended to the other modes, which appear vector-like 
in unbroken symmetries at low energy.
Their mass terms could be generated by D-brane instanton effects 
and eigenvalues of their masses may have a large hierarchy.

It is important to study the D-brane instanton induced $\mu$-term 
in intersecting D-brane models on orbifolds, too.
There are two types of cycles on the orbifolds.
One is a rigid cycle, which runs through the orbifold fixed points and 
the other is away from the fixed points\cite{Blumenhagen:2005tn}.
Then, in orbifold models, D-brane instantons can be classified into two types.
One is D-brane instanton which does not pass fixed points, and the other is that passes a fixed point. 
Their intersection numbers are different from each other.
Our results would be applicable in orbifold models, if 
the E-brane which does not pass fixed points can induce the $\mu$-term.
Then, we could realize a large hierarchy between the smallest and largest eigenvalues.
On the other hand, 
if the $\mu$-term is induced by the E-brane fixed at one of the orbifold  fixed points, 
there are no position moduli and our results are invalid.
In this case, we can not expect a hierarchical structure would appear 
because the ${\bf Z}_{n}$ translation symmetry is perfectly broken.
However, other typical structures might appear
because the fixed points of the orbifold are  special points, 
and another symmetry may appear.
Such a study would be interesting but it is beyond our scope and 
we would study elsewhere.


\subsection*{Acknowledgement}

H.A. was supported in
part by the Grant-in-Aid for Scientific Research No.~25800158 from the Ministry of Education,
Culture, Sports, Science and Technology (MEXT) in Japan. T.K. was supported in part by
the Grant-in-Aid for Scientific Research No.~25400252 from the MEXT in Japan.

\end{document}